\documentclass[journal=jacsat,manuscript=article]{achemso}

\usepackage[version=3]{mhchem} 

\author{Andrey Lyalin}
\email{lyalin@mail.sci.hokudai.ac.jp}
\altaffiliation{On leave from: Institute of Physics, St Petersburg State University, 
198504 St Petersburg, Petrodvorez, Russia}
\author{Tetsuya Taketsugu}
\affiliation[Hokkaido University]{Division of Chemistry, Graduate School of Science, 
Hokkaido University, Sapporo 060-0810, Japan}

\title{Adsorption of Ethylene on Neutral, Anionic and Cationic Gold Clusters}

\begin{document}

\begin{abstract}
The adsorption of ethylene molecule 
on neutral, anionic and cationic gold clusters consisting of up to 10 atoms 
has been investigated using density-functional theory.
It is demonstrated that \ce{C2H4} can be adsorbed
on small gold clusters in two different configurations,
corresponding to the $\pi$- and di-$\sigma$-bonded species.
Adsorption in the $\pi$-bonded mode dominates over the di-$\sigma$ 
mode over all considered cluster sizes $n$, with the exception of the 
neutral \ce{C2H4-Au5} system.
A striking difference is found in the size-dependence of the 
adsorption energy of \ce{C2H4} bonded to the neutral gold clusters in the 
$\pi$ and di-$\sigma$ configurations.
The important role of the electronic shell effects in the di-$\sigma$ mode 
of ethylene adsorption on neutral gold clusters is demonstrated. 
It is shown that the interaction of \ce{C2H4} with small gold clusters 
strongly depends on their charge.
The typical shift in the vibrational frequencies of \ce{C2H4}
adsorbed in the $\pi$- and the di-$\sigma$ configurations
gives a guidance to experimentally distinguish between the two modes of adsorption.
\end{abstract}

\section{Introduction}

The adsorption of unsaturated hydrocarbons on transition metal surfaces has been 
studied extensively in order to understand the nature of 
hydrocarbon -- metal interaction and chemical processes 
on solid surfaces; see, e.g., Refs. 
\cite{Astruc07,Masel96,Sheppard96,Sheppard98,Zaera95,Mrozek01}  and 
references therein.
The most significant attention was paid to 
investigation of ethylene adsorption,
\cite{Baerends72,Stuve85,Stuve85a,Cassuto90,Zinola98,Ge99,Watwe01,Itoh02,Paulus05,Kokalj06,Calaza07}
because it is the simplest alkene containing an isolated carbon--carbon 
double bond. Hence it can be treated as a prototype to study the interaction and reactivity
of different alkenes on metal surfaces. Moreover, the ethylene epoxidation is one of 
the most important processes in the chemical industry, because the product of such a reaction --
ethylene oxide is widely used in various applications;
see, e.g., Refs. \cite{vanSanten87,Barteau82,Torres05,Torres06,Astruc07,Masel96} 
and references therein.

Ethylene can adsorb on metal surfaces in two different 
configurations.\cite{Stuve85,Stuve85a,Zinola98,Astruc07} 
The first one is the $\pi$ mode where one metal atom on the surface
is involved in the adsorption of ethylene via a $\pi$-bonding.
The second one is the di-$\sigma$-bonded mode, when two metal atoms are involved
in the adsorption via a $\sigma$ bonding. 
It was found that the di-$\sigma$ mode of adsorption is 
characterized by the increasingly important role of $sp^3$ 
hybridization in ethylene, while $sp^2$ hybridization (typical for free molecules)
remains unchanged for $\pi$-bonded species.\cite{Stuve85,Stuve85a,Zinola98}

The bonding of ethylene with transition metals involves electron transfer from 
the filled bonding $\pi$ orbital of the ethylene to the 
metal, alongside a back-donation from the d-orbital of transition metal to the 
empty $\pi^*$ anti-bonding orbital of ethylene in accordance with the
Dewar-Chatt-Duncanson model \cite{Chatt53}.  
Thus the appearance of the $sp^3$ character of hybridization in 
the di-$\sigma$-bonded ethylene can be explained by 
the increasing role of the electron back-donation from the metal to the  
$\pi^*$ anti-bonding orbital of ethylene.
Hence the di-$\sigma$-bonded ethylene is activated more strongly in comparison with 
the $\pi$-bonded one, thus it can be more reactive.\cite{Stuve85,Stuve85a}
Therefore, an understanding of the specific mechanisms of ethylene adsorption on metal 
surfaces is important in order to gain a better insight of the catalytic processes and 
reactivity of adsorbed hydrocarbons.

Surprisingly, not many works have been devoted to the investigation 
of ethylene adsorption on metal clusters and nanoparticles, 
despite the fact that chemical and physical properties of matter 
at nanoscale are very different from those of the corresponding 
bulk solids. These properties are often controlled by quantum size effects and  
significantly depend on the size and structure of atomic clusters.\cite{Nanocat07}
 
A remarkable example is gold.
It is well known that gold in its bulk form is a catalytically inactive and inert metal. 
However gold at nanoscale manifests extraordinary catalytic activity 
which increases with a decrease in the cluster size of up to 1-5 nm 
\cite{Haruta87,Haruta97,Tsunoyama05,Turner08}.
Moreover, recent studies demonstrate that catalytic activity of gold clusters adsorbed on 
an iron oxide support correlates with the presence 
of very small clusters of $\sim$ 10 atoms \cite{Herzing08}.
It was also reported that gold clusters with the number of atoms 
$n = 3 - 11$ possess extraordinarily high electrocatalytic activity toward the \ce{O2} 
reduction reaction in acid solutions.\cite{Maria08}
The origin of such size-dependent catalytic activity of gold remains highly 
debated and has yet to be fully understood.
A comprehensive survey of the field can be found in review papers and books; 
see, e.g., Refs. \cite{Nanocat07,Pyykko08,Coquet08,Hakkinen08,Hvolbaek07,Johnson09}.
It was demonstrated that the catalytic activity of gold 
clusters depends on the type of the support material, the 
presence of defects in the material (e.g. F-center defects), 
the environment or special additives;
see, e.g., Refs. \cite{Haruta97,Sanchez99,Landman07,Coquet08,Tsukuda09} 
and references therein. 
However, experiments indicate that gold clusters 
deposited on an inert support 
can efficiently catalyze the process of styrene oxidation 
by dioxygen \cite{Turner08}. Hence, the origin of catalytic activity derives from the
cluster itself and therefore even free clusters can be effective catalysts.

Currently, a large variety of catalytic reactions on gold clusters 
has been studied experimentally. This includes the processes of catalytic oxidation  
of carbon monoxide, as well as more complex reactions such as alcohol oxidation, 
the direct synthesis of hydrogen peroxide and alkene 
epoxidation.\cite{Turner08,Hughes05,Tsunoyama05,Landon02}

However, theoretical studies of the gold nanocatalysis 
have mainly focused on the investigation of adsorption and catalytic reaction of 
\ce{O2} and \ce{CO}.\cite{Coquet08,Landman07} 
Nonetheless, the theoretical study of the interaction of alkenes with 
gold clusters has been relatively unexplored 
with the exception of works on the binding of propene on gold and 
mixed gold-silver clusters\cite{Metiu04,Metiu04auag},
propene epoxidation\cite{Lee09} and our recent work 
on cooperative adsorption of oxygen and ethylene on small 
gold clusters\cite{Lyalin09}. 

In this paper we report the results of a systematic theoretical study 
of the adsorption of \ce{C2H4} on free neutral, anionic and cationic 
gold clusters consisting of up to 10 atoms.
We find that \ce{C2H4} adsorbs on small gold clusters in two different configurations,
corresponding to the $\pi$- and di-$\sigma$-bonded species.
Adsorption in the $\pi$-bonded mode dominates over di-$\sigma$ 
mode over all considered cluster sizes $n$ with the exception of the 
neutral \ce{C2H4-Au5} system, where the 
di-$\sigma$-bonded configuration is energetically more favorable. 
We find a striking difference in the size dependence of the 
adsorption energy of \ce{C2H4} bonded to neutral gold clusters in the 
$\pi$ and di-$\sigma$ configurations.  
Thus, the adsorption energy of the $\pi$-bonded \ce{C2H4} develops 
non-monotonically as a function of cluster size with a local minima at $n$=6. 
The adsorption energy, calculated for the di-$\sigma$-bonded \ce{C2H4} exhibits 
pronounced odd-even oscillations, showing the importance of the electronic 
shell effects in the di-$\sigma$ mode of ethylene adsorption on gold clusters. 

We also demonstrate that the interaction of \ce{C2H4} with small gold clusters 
strongly depends on their charge. The excess of a positive or a negative charge on 
the gold cluster can change the balance between electron donation and 
back-donation processes. 
Hence ethylene adsorption and reactivity  
can be manipulated by the cluster charge.
This effect can be particularly important for understanding the mechanisms of catalytic
activation of ethylene adsorbed on small gold clusters.

\section{Theoretical Methods}

The calculations are carried out using density-functional theory (DFT). 
The hybrid Becke-type three-parameter 
exchange density functional paired with the gradient-corrected 
Perdew-Wang 91 correlation functional (B3PW91) \cite{PerWan} is used.
The choice of the B3PW91 functional is consistent with our previous work 
on cooperative adsorption and catalytic oxidation of ethylene on 
small gold clusters.\cite{Lyalin09} 
The obtained theoretical data on dissociation energies and bonding in 
\ce{Au2} (1.904 eV, 2.546  \AA), and 
\ce{C2H4} (7.596 eV, 1.324 \AA) are in a good agreement 
with those of earlier experimental studies, 
\ce{Au2} (2.30 eV,  2.472 \AA) \cite{Herzberg79}, and 
\ce{C2H4} (7.76 eV,  1.339 \AA) \cite{Carter88}.
The standard LANL2DZ basis set of primitive Gaussians is used to 
expand the gold cluster orbitals formed by the $5s^2 5p^6 5d^{10} 6s^1$ 
outer electrons of Au (19 electrons per atom). 
The remaining 60 core electrons of the Au atom are represented 
by the Hay-Wadt effective core potential accounting for relativistic effects 
\cite{Wad-Hay}.  For carbon and hydrogen, the aug-cc-pVTZ basis set\cite{Dunning89} is employed. 
Calculations have been carried out with the use of
the Gaussian 03 code \cite{Gaussian}.

\begin{figure}[htbp]
\includegraphics[scale=0.70,clip]{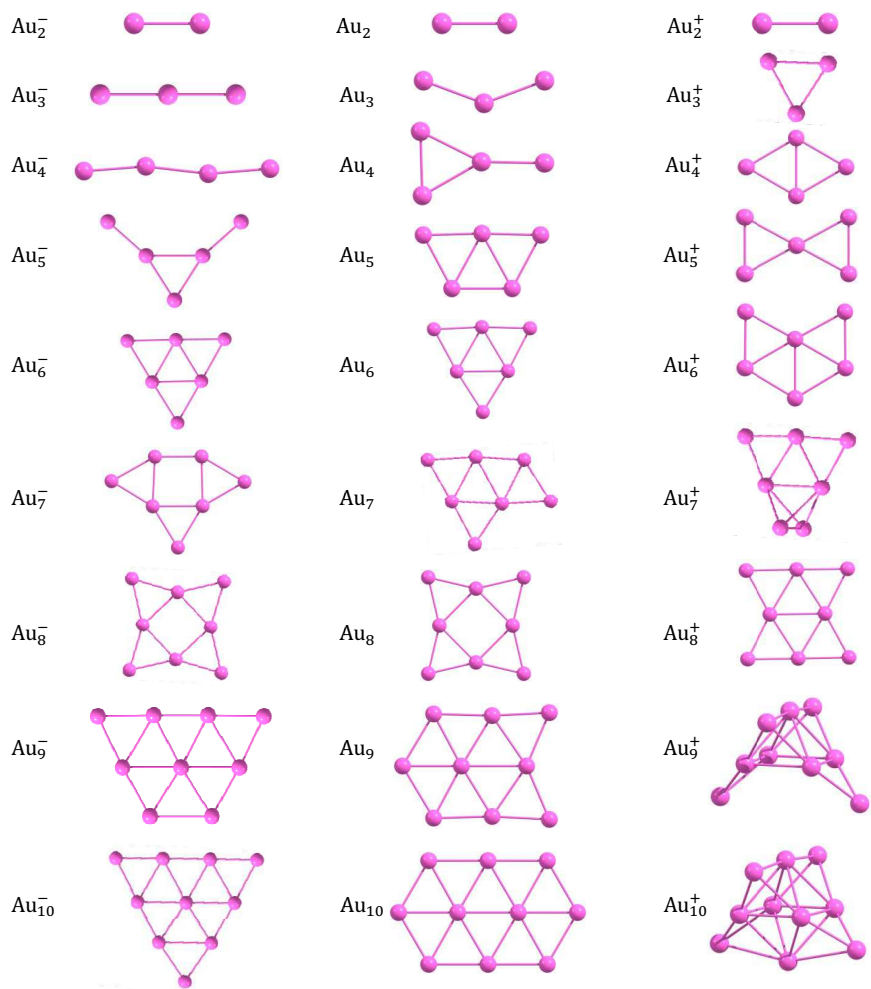}
\caption{Optimized geometries of the most stable anionic, \ce{Au_n^-}, 
neutral, \ce{Au_n}, and cationic, \ce{Au_n^+}, gold clusters 
calculated within the size range n= 1, ., 10 in the B3PW91/LanL2DZ approximation.}
\label{fig:geom_gold}
\end{figure} 

The structural properties of neutral, anionic and cationic gold clusters have been the subject 
of numerous theoretical investigations; see, e.g., Refs. \cite{Pyykko08,Coquet08,Hakkinen08} 
for a review.  
In the present work the cluster geometries have been determined with the use of the 
cluster fusion algorithm\cite{LJ_PRL03} which belongs to the class 
of genetic global optimization methods \cite{Goldberg89,Michalewicz96}. 
We have successfully used a similar 
approach to find the optimized geometries of various types of atomic clusters  
\cite{struct_Mg,struct_La,struct_Sr}.
The optimized structures of the gold clusters \ce{Au_n}, \ce{Au_n^-}, and \ce{Au_n^+} 
($n = 1 - 10$) obtained in the present work 
are presented in \ref{fig:geom_gold}. These structures 
are in a good agreement with those reported in previous theoretical studies; see, e.g., 
Refs. \cite{Ding04,Fernandez04,Walker05,Xiao06,Hakkinen08}. 

In order to obtain the most stable configuration of \ce{C2H4}
adsorbed on neutral, anionic and cationic gold clusters,
we have created a large number of starting geometries by adding \ce{C2H4} molecule
in different positions (up to 30) on the surface of the most stable cluster 
and up to eight isomer structures of the corresponding  \ce{Au_n}, \ce{Au_n^-}, and \ce{Au_n^+}
clusters. The starting structures have been optimized further without any geometry constraints.

\section{Theoretical results}
\label{results}

\subsection{Geometry optimization for ethylene adsorbed on neutral, cationic and anionic 
gold clusters}
\label{geom_opt}
         
\begin{figure}[htbp]
\includegraphics[scale=0.70,clip]{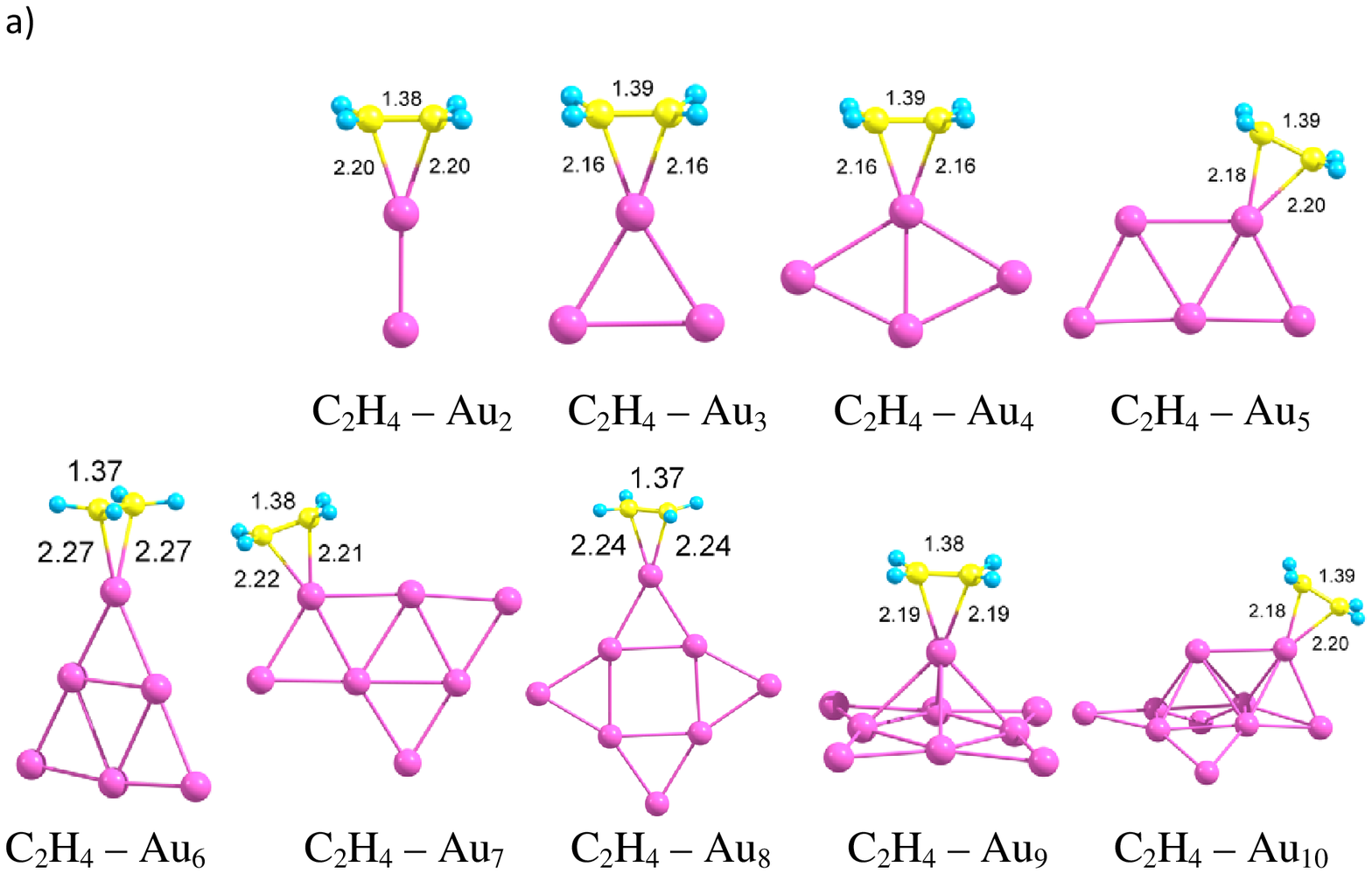}
\includegraphics[scale=0.70,clip]{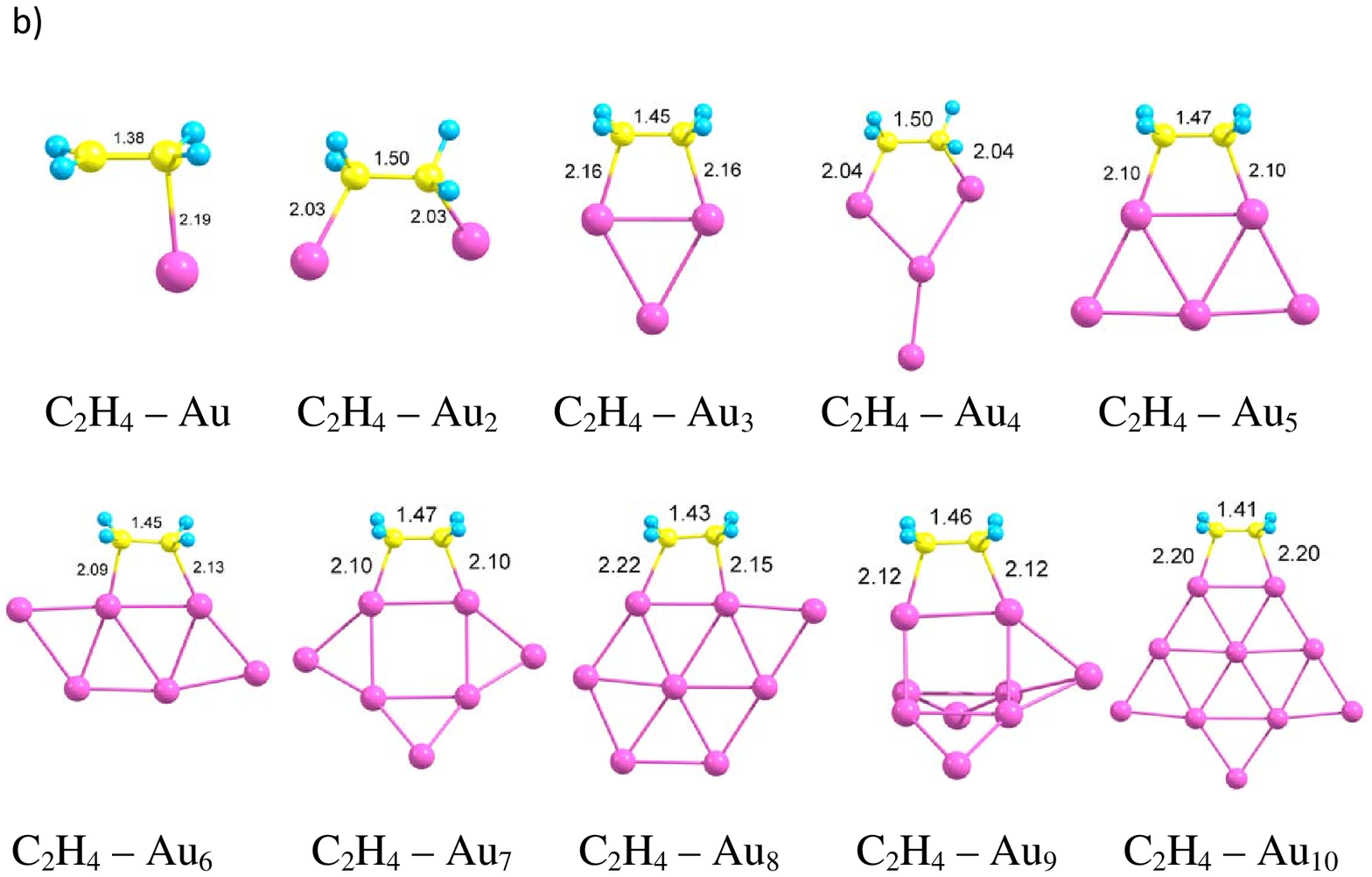}
\caption{Optimized geometries of neutral \ce{C2H4-Au_n} clusters:
(a) $\pi$-mode of adsorption;
(b) di-$\sigma$-mode of adsorption 
(in the case of a single Au atom only one $\sigma$ bond is formed).
The C -- C and C -- Au interatomic distances are given in angstroms.}
\label{fig:geom_neutral}
\end{figure}

\begin{figure}[htbp]
\includegraphics[scale=0.70,clip]{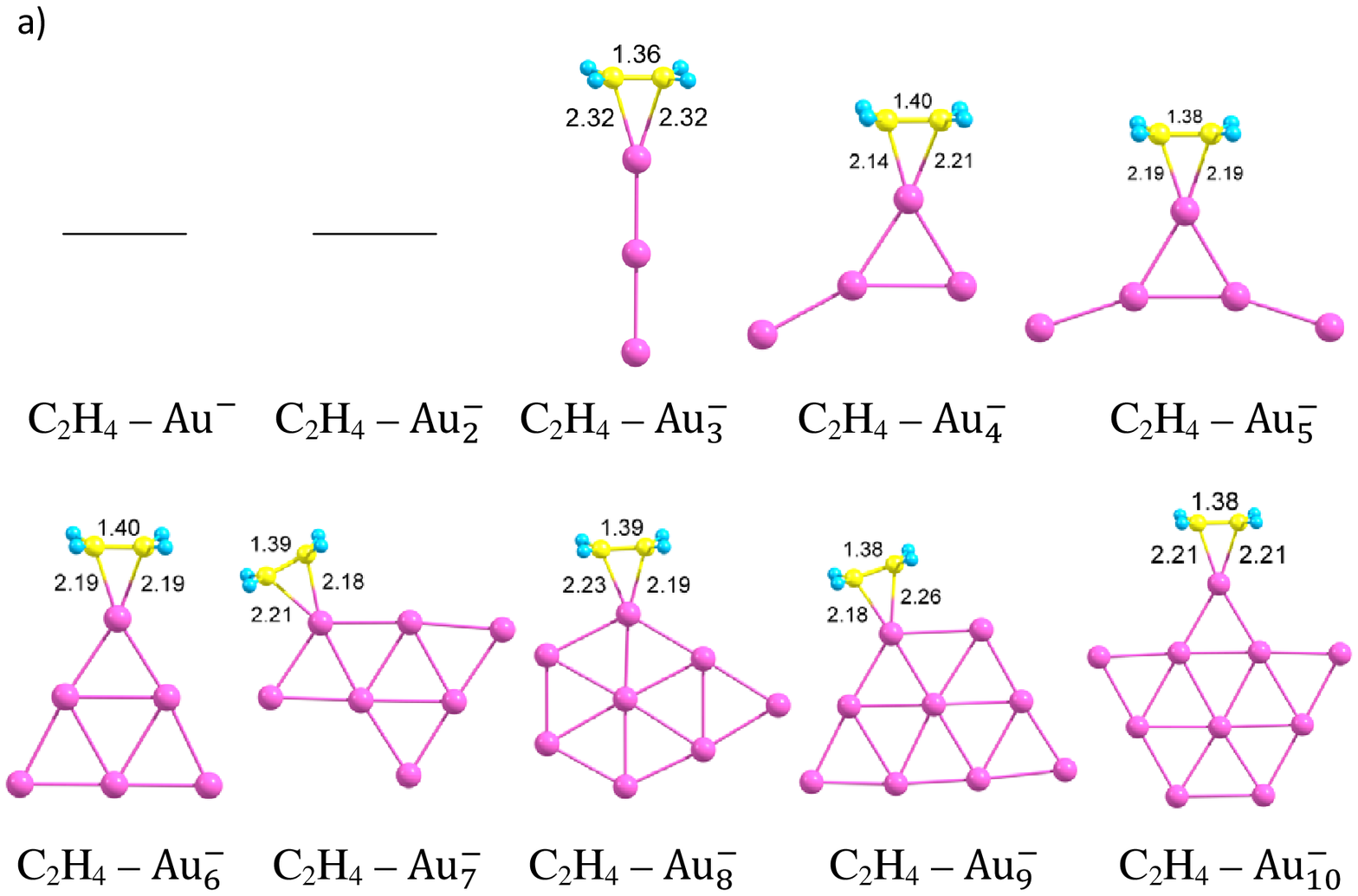}
\includegraphics[scale=0.70,clip]{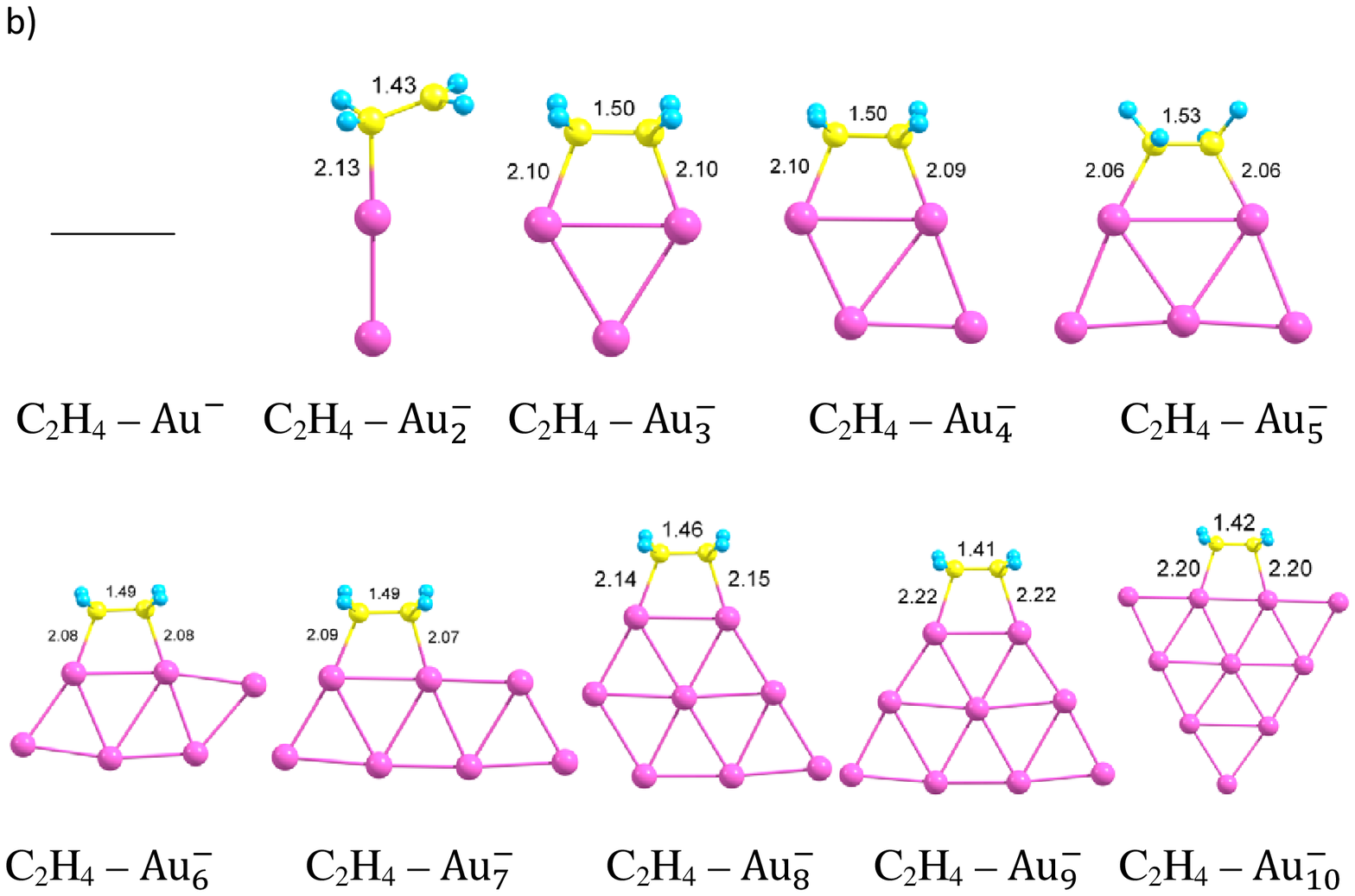}
\caption{Same as in \protect{\ref{fig:geom_neutral}}, 
but for anionic \ce{C2H4-Au_n^-} clusters.
In the case of a \ce{Au2^-} dimer only one $\sigma$ bond is formed.}
\label{fig:geom_anion}
\end{figure}

\begin{figure}[htbp]
\includegraphics[scale=0.70,clip]{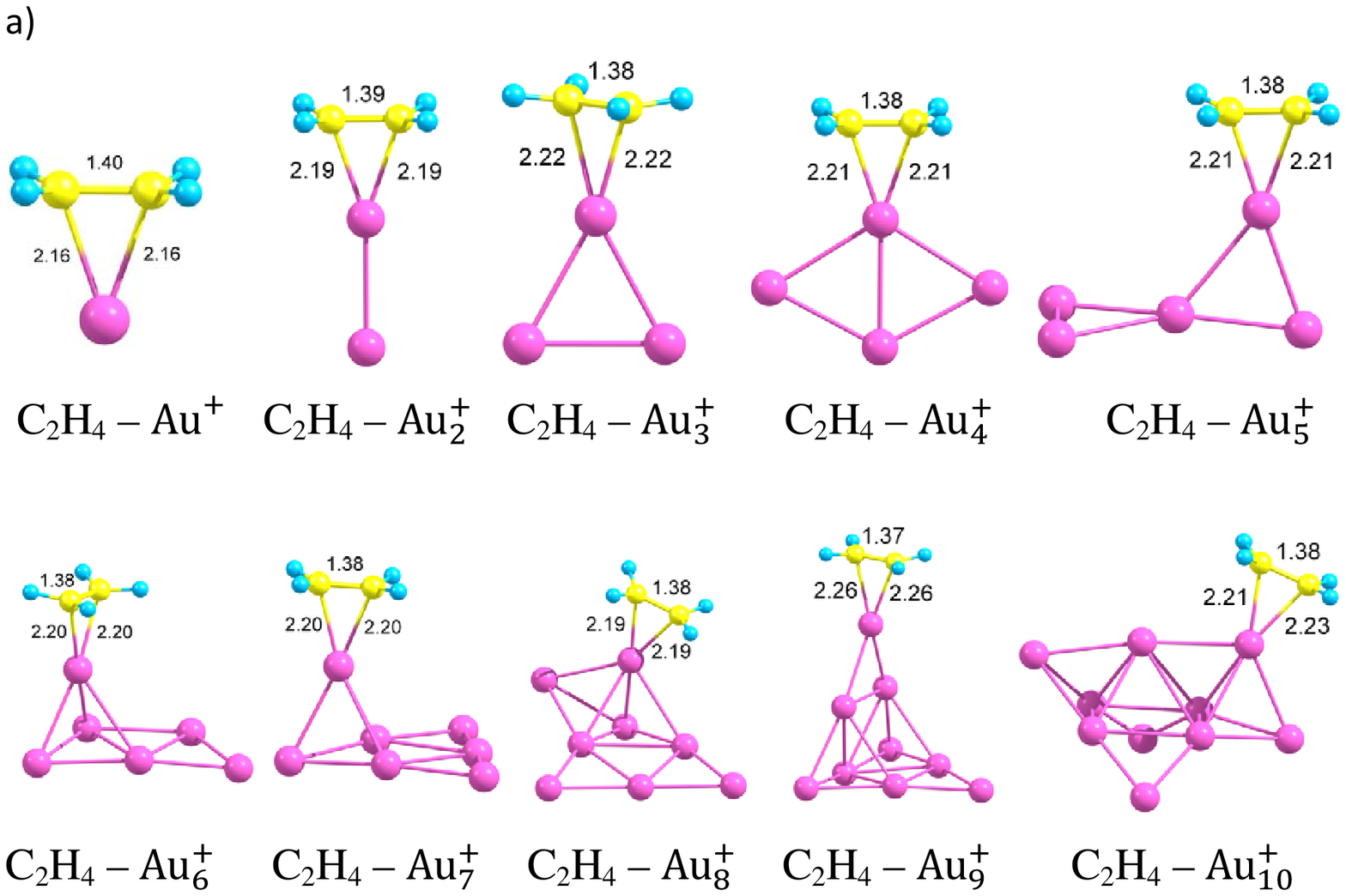}
\includegraphics[scale=0.70,clip]{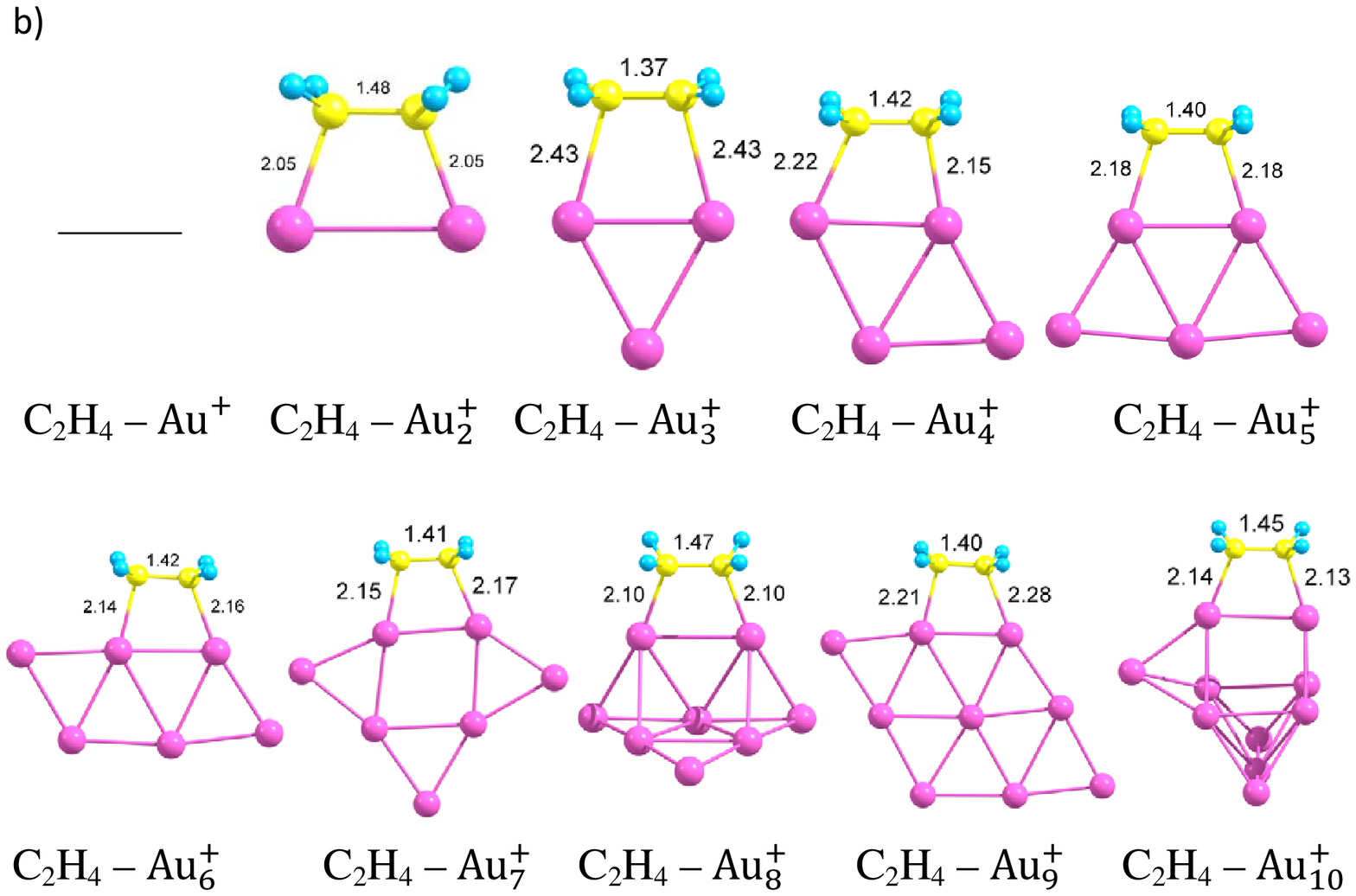}
\caption{Same as in \protect{\ref{fig:geom_neutral}}, 
but for cationic \ce{C2H4-Au_n^+} clusters.
}
\label{fig:geom_cation}
\end{figure}

The results of the cluster geometry optimization 
for neutral \ce{C2H4-Au_n}, 
anionic \ce{C2H4-Au_n^-}
and cationic \ce{C2H4-Au_n^+} clusters 
within the size range $n \le 10 $
are shown in \ref{fig:geom_neutral}, \ref{fig:geom_anion} and \ref{fig:geom_cation},  
respectively.

Gold clusters with adsorbed ethylene possess various isomer forms whose number
grows dramatically with cluster size. 
Among the huge variety of different isomers, one can distinguish  
two specific adsorption modes that correspond to the
$\pi$- and di-$\sigma$ configurations of the adsorbate. 
The appearance of two different modes of ethylene adsorption on small gold clusters is 
similar to that reported for ethylene adsorption on Au(100) and Au(111) surfaces\cite{Zinola98}. 
However in the case of adsorption on the bulk, there 
are only two possible geometrical configurations of ethylene (we do not consider here 
$\mu$-bridging species\cite{Zinola98}, 
because ethylene adsorption in such a configuration is not possible 
in the case of small clusters). Moreover, there is no structural changes 
of the bulk surface due to adsorption of ethylene. 
In the case of ethylene adsorption on small gold clusters, 
each of the $\pi$- and di-$\sigma$ configurations possess a variety of structural isomers. 
These isomers correspond to the adsorption in nonequivalent 
positions on the surface of clusters. 
In \ref{fig:geom_neutral}, \ref{fig:geom_anion} and \ref{fig:geom_cation}
we present the most stable  structures of $\pi$- and di-$\sigma$-bonded 
ethylene adsorbed on small gold clusters.
The C -- C and C -- Au interatomic distances are given in angstroms.


The interaction of gold clusters with ethylene 
results in an electron donation from the \ce{C2H4} 
to the gold cluster alongside a back-donation from 
the cluster to the adsorbate in accordance with the
Dewar-Chatt-Duncanson model \cite{Chatt53}.
The rearrangement of the electronic structure and
electron transfers upon adsorption lead to a dramatic 
change in cluster geometry. The interplay of electronic 
and geometry shell effects, also influence the cluster structure \cite{struct_Sr}.

Free neutral gold clusters possess planar, two-dimensional (2D) 
structures when the cluster size is up to 13 Au atoms.\cite{Hakkinen03,Xiao06} 
In \ref{fig:geom_gold}, we present the most stable structures obtained for \ce{Au}$_n$,  
\ce{Au_n^+} and  \ce{Au_n^-} clusters ($n$ = 1, ..., 10) 
within the B3PW91/LANL2DZ DFT method. 
\ref{fig:geom_neutral} demonstrates that adsorption  of \ce{C2H4} 
in the $\pi$ configuration leads to a prevalence of three-dimensional (3D) structures 
of gold clusters for $n \ge 9$; the structures of small \ce{Au3} and \ce{Au4} clusters 
transform to that which are energetically favorable for free cations 
\ce{Au_3^+} and \ce{Au_4^+}, respectively.
For $n < 9$, ethylene favorably adsorbs at the cluster edge 
where the carbon-carbon bond lay in the plane of the cluster with the exception of the 
\ce{C2H4-Au_6} system, where the carbon-carbon bond is located perpendicular 
to the cluster plane (see \ref{fig:geom_neutral}). 

Adsorption of \ce{C2H4} in the di-$\sigma$ configuration results in a considerable
rearrangement in the structure of neutral gold clusters.
Thus, the geometry of \ce{Au3}, and \ce{Au6} -- \ce{Au10} clusters changes upon 
\ce{C2H4} adsorption. It is known that gold clusters possess a large number of 
energetically close-lying structural isomers.\cite{Pyykko08,Hakkinen08} 
Interaction with adsorbate can easily alter their energy order.  
It is important to note, that interaction with adsorbate can be responsible for 
structural rearrangements even for the relatively large clusters. Thus, 
recent DFT calculations demonstrate that the morphology of \ce{Au79} nanoparticle 
can be transformed by interactions with a CO atmosphere.\cite{McKenna07}

It is not surprising that the structures of positively and negatively charged 
ionic gold clusters change upon ethylene adsorption. 
\ref{fig:geom_anion} demonstrates that although anionic gold 
clusters \ce{Au_n^-} with $n$ = 1, ..., 10 remain planar upon 
ethylene adsorption (both for the $\pi$ and di-$\sigma$ adsorption configurations),
they manifest considerable structural rearrangements.
Thus, gold cluster anions alter their structure upon ethylene adsorption 
in the $\pi$-configuration for $n$ = 4, 7, 8, and 10.
In the case of di-$\sigma$ mode of ethylene adsorption, 
structural rearrangements occur for gold cluster anions with $n$ = 3 -- 8.  
A carbon-carbon bond of the adsorbed ethylene lie in a cluster plane 
for all structures presented in \ref{fig:geom_anion}. 

\ref{fig:geom_cation} shows that in the case of ethylene adsorption 
on gold cluster cations, \ce{Au_n^+}, the resulting cluster systems 
\ce{C2H4-Au_n^+} exhibit 3D structures for
$n$= 5 -- 10 in the case of the $\pi$ mode of ethylene adsorption; and $n$ = 8 and 10 in the case 
of the di-$\sigma$ mode.  In both cases of the $\pi$ and di-$\sigma$ modes of ethylene adsorption, 
the structural rearrangements in gold cluster cations occur for $n \ge$ 5. 

Such a severe change in the morphology of gold clusters due to the interaction with ethylene 
makes the procedure of structural optimization rather complicated. 
In order to get energetically favorable structures of 
\ce{C2H4-Au_n}, \ce{C2H4-Au_n^-}, and \ce{C2H4-Au_n^+} clusters, it is 
necessary to consider adsorption of ethylene not only 
on the most stable structures of the corresponding 
\ce{Au_n}, \ce{Au_n^-}, and \ce{Au_n^+} clusters, but also on a large 
number of their energetically less favorable isomer states.
In the present work we have taken into account up to 
8 isomer structures of the host gold clusters. 
This insures that we do not miss energetically favorable structures with considerable 
change in cluster geometry due to \ce{C2H4} adsorption.

\subsection{Energetics of ethylene adsorption on neutral, anionic and cationic gold clusters}
\label{Adsorption_energy}

\begin{figure}[htb]
\includegraphics[scale=1.2,clip]{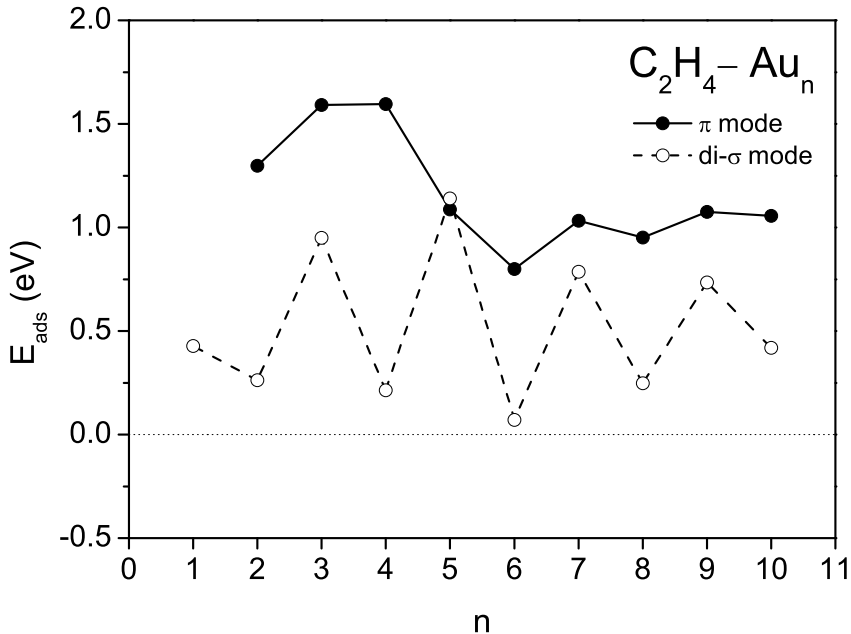}
\caption{Molecular adsorption energy, $E_{\rm ads}$, 
calculated for C$_2$H$_4$ 
adsorbed in $\pi$ (solid line) and di-$\sigma$ (dashed line) configurations 
on the neutral \ce{Au_n} clusters with $1 \le n \le 10$.
Note, that in the case of a single Au atom only one $\sigma$ bond is formed.}
\label{fig:Eads_neutral}
\end{figure}

We first study the energetics of \ce{C2H4} adsorption on neutral gold clusters 
Au$_n$ with $1 \le n \le 10$.
The spin states of the optimized \ce{C2H4-Au_n} structures 
are doublet and singlet for odd and even $n$, respectively.
\ref{fig:Eads_neutral} shows the evolution of the 
molecular adsorption energy, $E_{\rm ads}$, calculated for the most bounded
$\pi$ and di-$\sigma$ configurations of adsorbed \ce{C2H4}
as a function of cluster size $n$.
The adsorption energy is defined as 
\begin{equation}
E_{\rm ads} =  E_{tot}({\rm Au}_n) + E_{tot}({\rm C}_2{\rm H}_4) - 
               E_{tot}({\rm Au}_n-{\rm C}_2{\rm H}_4), 
\end{equation}
\noindent where $E_{tot}(M)$ denotes the total energy of the most stable structure 
of the molecule (cluster) "$M$".

In our previous work, we have studied adsorption of ethylene on small neutral gold clusters 
in the $\pi$-bonded configuration.\cite{Lyalin09}
Here, we demonstrate that \ce{C2H4}
can be adsorbed on small gold clusters in two different configurations,
corresponding to the $\pi$- and di-$\sigma$-bonded species.
\ref{fig:Eads_neutral} shows that the $\pi$-bonded mode of \ce{C2H4} adsorption is energetically favorable
in the considered range of cluster sizes $n$, with the 
exception of a single Au atom 
which interacts with one carbon atom in \ce{C2H4}, forming a $\sigma$ bond, 
and the \ce{Au5} cluster, where  the di-$\sigma$-bonded configuration of 
\ce{C2H4-Au_n} is 0.05 eV more bounded compared with the $\pi$-bonded one.  

It is seen from \ref{fig:Eads_neutral} that the adsorption energy of 
the $\pi$-bonded \ce{C2H4} on \ce{Au_n} clusters exhibit a maximum 
at $n=$ 3 and 4 followed by a fast drop with a minimum at $n=$ 6.
For $6 \le n \le 10$ the adsorption energy of $\pi$-bonded \ce{C2H4}
shows a weak oscillatory  behavior as a function of $n$.

The $n$-evolution of the adsorption energy calculated for 
the di-$\sigma$-bonded \ce{C2H4} is considerably different 
from that of the $\pi$-bonded \ce{C2H4}.
\ref{fig:Eads_neutral} demonstrates that the adsorption energy of 
the di-$\sigma$-bonded \ce{C2H4} has an odd-even oscillatory behavior 
with a local maxima for odd numbers of Au atoms $n=$ 3, 5, 7, and 9.
For clusters with an even number of $n$, the adsorption energy $E_{\rm ads}$ 
of the di-$\sigma$-bonded \ce{C2H4} is small. 
Hence, \ce{C2H4} readily adsorbs in the di-$\sigma$ configuration---but  
only for neutral gold clusters with an odd number of Au atoms. 

It is interesting to note that in the case of \ce{C2H4} adsorption on a Au(111) surface,
the di-$\sigma$ mode ($E_{\rm ads}^{di-\sigma}$ = 0.6502 eV) is
energetically more favorable in comparison to the $\pi$ mode 
($E_{\rm ads}^{\pi}$ = 0.1545 eV) \cite{Zinola98}.  
A similar effect is observed for \ce{C2H4} adsorbed on a Au(100) surface where
theoretical values of $E_{\rm ads}^{di-\sigma}$ = 0.7790 eV and 
$E_{\rm ads}^{\pi}$ = 0.2456 eV have been reported \cite{Zinola98}.
It is known that the type of \ce{C2H4} bonding depends upon the surface structure. 
It has been found that the $\pi$-bonded mode dominates in the adsorption of  
low coordinated atoms and step sites \cite{Rioux08}.
In the case of small metal clusters, almost all atoms are essentially 
surface atoms that are undercoordinated as compared to the bulk.\cite{Landman09} 
Thus, the $\pi$-bonded configuration of the adsorbed \ce{C2H4} dominates for 
small gold clusters.  As the number of low coordinated atoms  
decreases with increase in cluster size, we suppose that 
the di-$\sigma$ mode will become dominant for large gold clusters.

\begin{figure}[htb]
\includegraphics[scale=1.2,clip]{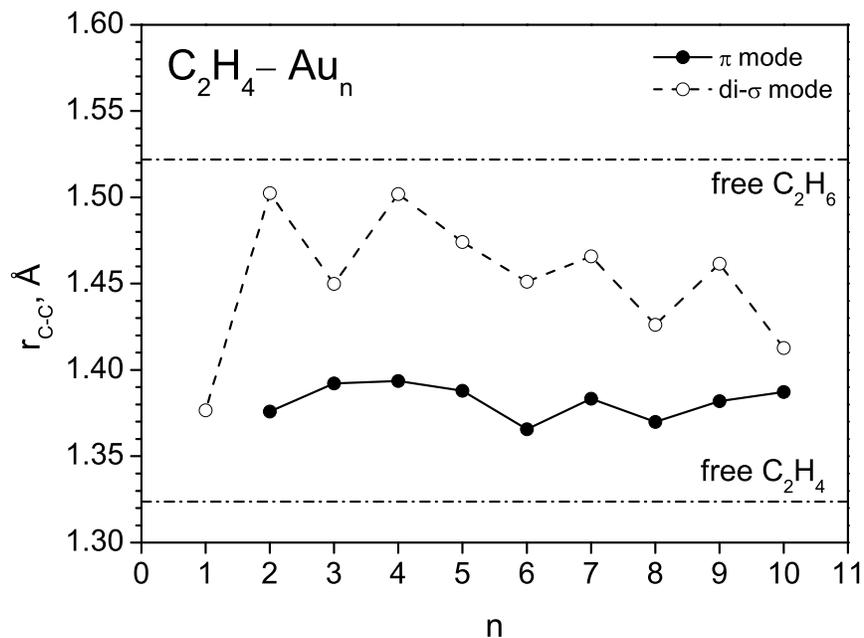}
\caption{The C--C bond distance, $r_{C-C}$, in \ce{C2H4} 
adsorbed in the $\pi$ (solid line) and di-$\sigma$ (dashed line) configurations 
on the neutral \ce{Au_n} clusters with $1 \le n \le 10$.
The dashed-dotted lines denote the equilibrium C--C bond distances for free 
\ce{C2H4} and \ce{C2H6}.
Note, that in the case of a single Au atom only one $\sigma$ bond is formed.}
\label{fig:R_neutral}
\end{figure} 

\begin{figure}[htb]
\includegraphics[scale=1.2,clip]{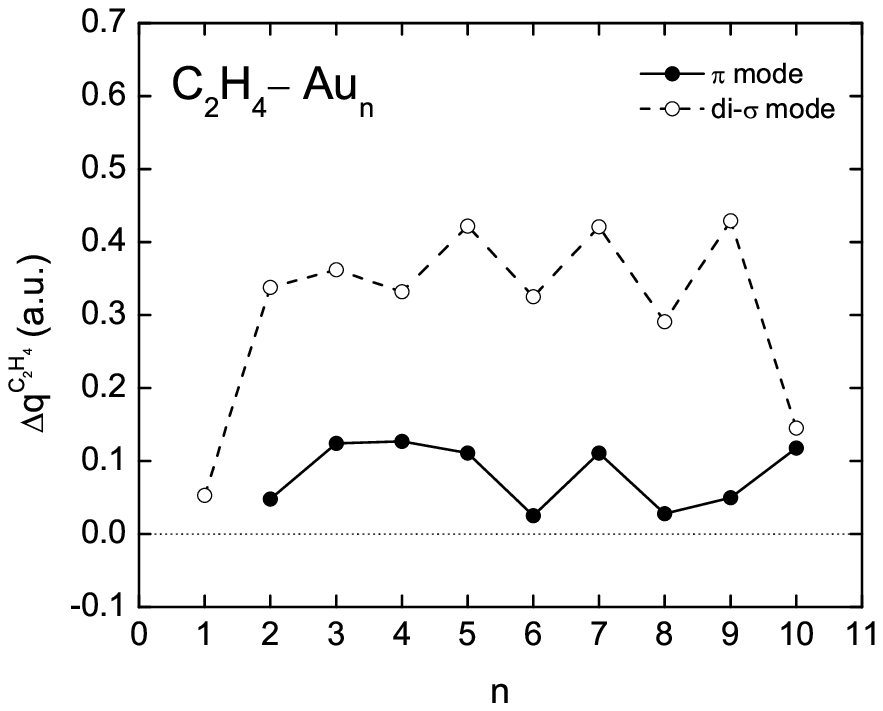}
\caption{The electron charge transfer, $\Delta q^{{\rm C}_2{\rm H}_4}$, to \ce{C2H4}
adsorbed in the $\pi$ (solid line) and di-$\sigma$ (dashed line) configurations 
on the neutral \ce{Au_n} clusters with $1 \le n \le 10$.
Note, that in the case of a single Au atom only one $\sigma$ bond is formed.}
\label{fig:Charge_neutral}
\end{figure}

\ref{fig:R_neutral} and \ref{fig:Charge_neutral}  
present the $n$-evolution of the the C--C bond distance, $r_{C-C}$,
and the total charge transfer to the adsorbate, $\Delta q^{{\rm C}_2{\rm H}_4}$, 
obtained from the natural bond orbitals (NBO) population analysis\cite{NBO88}.
\ref{fig:R_neutral} shows that the C--C bond in the $\pi$-bonded \ce{C2H4} 
is enlarged up to 1.37--1.39 \AA{} in comparison with the case of a 
free \ce{C2H4} (1.32 \AA). The $r_{C-C}$ calculated for the $\pi$-bonded \ce{C2H4} 
depends slightly on the cluster size, evolving with $n$
similar to $E_{\rm ads}$. 
On the other hand, for the di-$\sigma$-bonded ethylene the C--C bond 
is enlarged up to 1.42--1.50 \AA{}, which is comparable with the length 
of a single C--C bond in ethane, \ce{C2H6}.
The $n$-dependence of $r_{C-C}$ for the di-$\sigma$-bonded \ce{C2H4} manifests an irregular 
behavior with maxima at $n=$ 2, 4, 7 and 9. 
An increase in the C--C bond length, as well as as noticeable change of the 
bending of the H atoms for di-$\sigma$-bonded \ce{C2H4} (see \ref{fig:geom_neutral}), 
can be explained by the increasingly important role of the $sp^3$ hybridization 
due to the electron donation from the gold cluster to the 
anti-bonding $\pi^*$ orbital of \ce{C2H4}.  Therefore, $r_{C-C}$ in the di-$\sigma$-bonded 
ethylene approaches its typical values for a single C--C bond in $sp^3$ hybridized \ce{C2H6}.
Hence the di-$\sigma$-bonded ethylene is activated strongly in comparison with 
the $\pi$-bonded one, thus it can be more reactive in catalytic processes.

The total charge transfer to the adsorbed \ce{C2H4}     
depends on the balance between the donation and back-donation processes.
\ref{fig:Charge_neutral}  demonstrates that in the case of 
the $\pi$ mode of adsorption the total charge transfer 
$\Delta q^{{\rm C}_2{\rm H}_4}$  is relatively small and 
evolves with $n$ similar to $r_{C-C}$. For the di-$\sigma$ mode of 
adsorption the back-donation from the gold  to the 
$\pi^*$ anti-bonding orbital of ethylene dominates over electron transfer 
from \ce{C2H4} to the cluster. The total charge transfer 
to \ce{C2H4} adsorbed in a di-$\sigma$ configuration is large and varies 
in the range of 0.29 - 0.43 a.u. for $n=$ 2 - 9. 
It is seen from 
\ref{fig:Charge_neutral} that $\Delta q^{{\rm C}_2{\rm H}_4}$ calculated 
for the di-$\sigma$-bonded \ce{C2H4}
exhibits a pronounced odd-even oscillations as a function of cluster size $n$ with the 
exception of $n=$1. Note, that for $n=$1 the charge transfer $\Delta q^{{\rm C}_2{\rm H}_4}$
is small because in this case only one \ce{Au-C} $\sigma$ bond is formed. 
Similar odd-even oscillations have been found in the $n$-dependence
of the adsorption energy of \ce{O2} on \ce{Au_n} clusters; see, e.g. 
\cite{Ding04,Fernandez05,Nanocat07} and references therein.
However, the adsorption energy of \ce{CO} does not present any 
odd-even effects\cite{Fernandez05}.

The appearance of the odd-even oscillations in $E_{\rm ads}$ 
and $\Delta q^{{\rm C}_2{\rm H}_4}$ 
for the di-$\sigma$ mode of adsorption is a result of 
the electronic shell effects\cite{Knight84,Ekardt84,Hakkinen08} 
in the gold clusters and can be described within the jellium model.
In spite of its simplicity, the jellium model can explain, 
at least on the  qualitative level, many physical properties and chemical reactivity of 
metal clusters.\cite{Hakkinen08,Khanna09,Jellium_IJMPE03}
Indeed, each Au atom has one $6s$ electron which 
delocalizes in the whole volume of the cluster. 
The delocalized electrons are moving in the 
field of a uniform positive charge background, which binds the valence electron cloud.
The gold clusters with an even number of atoms (valence electrons)
have closed electronic shell structure.
Hence these clusters are more stable and less reactive in comparison with  
open shell clusters possessing an odd number of atoms (valence electrons).

The prevalence of the electron transfer from the 
cluster to the adsorbate (back-donation process) is responsible for the promotion of the 
di-$\sigma$-mode of \ce{C2H4} adsorption.
Indeed, the di-$\sigma$ mode of \ce{C2H4} adsorption on small gold clusters with an 
odd number of atoms leads to the transfer of unpaired electron from the gold cluster to the 
ethylene anti-bonding $\pi^*$ orbital, causing the weakening of the C--C bond.
The neutral gold clusters 
with even number of valence electrons possess a closed electronic shell structure, 
hence electron transfer to the adsorbate and 
formation of the di-$\sigma$ bonded species is suppressed.

\begin{figure}[htb]
\includegraphics[scale=0.8,clip]{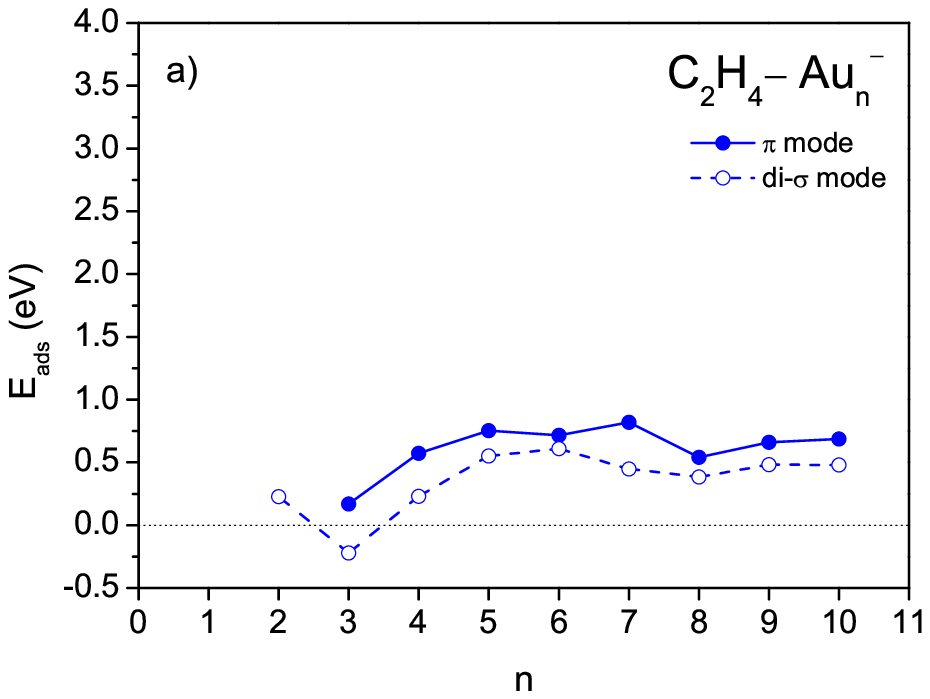}
\includegraphics[scale=0.8,clip]{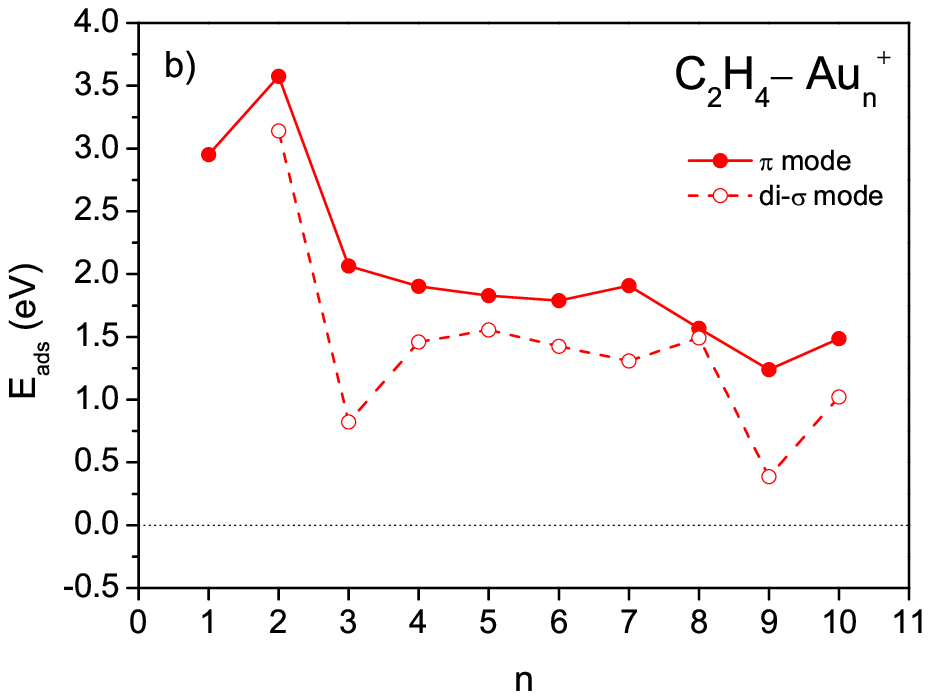}
\includegraphics[scale=0.8,clip]{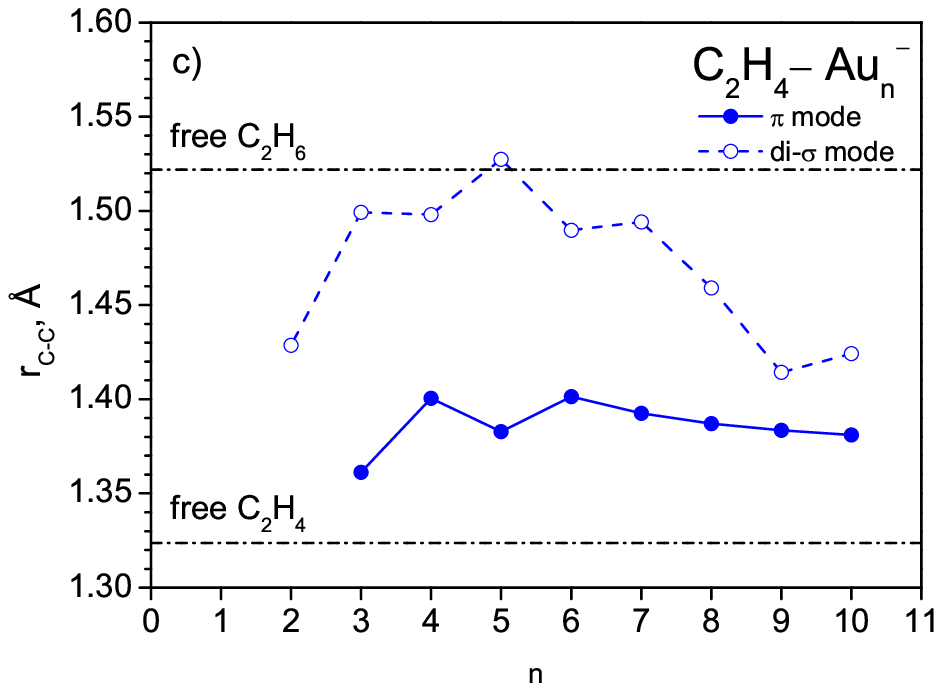}
\includegraphics[scale=0.8,clip]{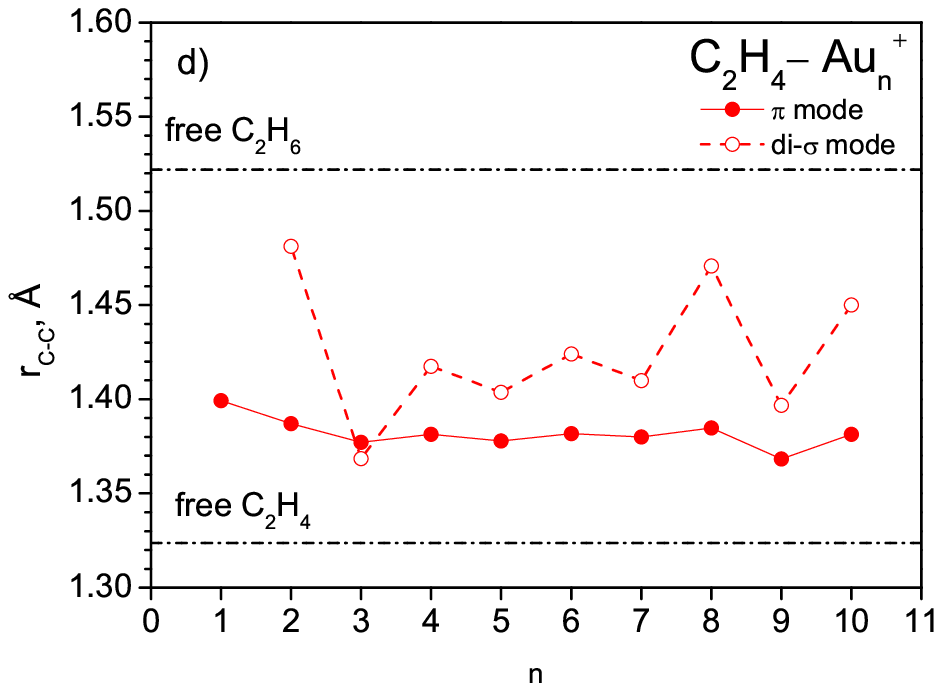}
\caption{Molecular adsorption energy, $E_{\rm ads}$, (upper row);
and C--C bond distances in adsorbed molecules, $r_{C-C}$, (lower row); 
calculated for \ce{C2H4} adsorbed on the anionic  \ce{Au_n^-}  (left column) 
and cationic \ce{Au_n^+} (right column) clusters with $1 \le n \le 10$.
Solid and dashed lines correspond to the $\pi$ and di-$\sigma$  
configurations of adsorbed \ce{C2H4}, respectively.
Dashed-dotted lines in lower row figures denote 
the equilibrium C--C bond distances for free \ce{C2H4} and \ce{C2H6}.}
\label{fig:ads_minus_plus}
\end{figure} 

\ref{fig:ads_minus_plus} demonstrates the evolution of 
molecular adsorption energy, $E_{\rm ads}$ and C--C bond distances, $r_{C-C}$,  
as a function of cluster size $n$ for \ce{C2H4} adsorbed 
on the anionic  \ce{Au_n^-}  (left column) and cationic \ce{Au_n^+} (right column) 
clusters with $1 \le n \le 10$.

The adsorption energy of \ce{C2H4} on a gold cluster depends on the 
balance between donation and back-donations processes, hence
it can be manipulated by the cluster charge. 
It is seen form \ref{fig:ads_minus_plus}a and \ref{fig:ads_minus_plus}b
that excess of the negative or the positive charges on gold clusters
strongly influence the molecular adsorption. 
Thus, the interaction of \ce{C2H4} with the negatively charged gold clusters \ce{Au_n^-} 
becomes weaker if compared with the neutrals; 
while excess of the positive charge results in strengthening of the \ce{C2H4} and \ce{Au_n^+} bond. 

\ref{fig:ads_minus_plus}a demonstrates that the $\pi$-mode of \ce{C2H4} adsorption is 
not stable for \ce{Au^-} and \ce{Au_{2}^{-}}.
The adsorption energy of the $\pi$-bonded \ce{C2H4} on the gold cluster anions
\ce{Au_{n}^{-}} is 0.17 eV in the case of a trimer \ce{Au_{3}^{-}}, 
and this increases with an increase in the cluster size: up to 0.82 eV at $n=$7 and dropping  
to 0.54 eV at $n=$8, while slowly increasing with an increase in the cluster size 
up to $n=$10. 

\ref{fig:ads_minus_plus}b shows that the size dependence of $E_{\rm ads}$ for 
$\pi$-bonded \ce{C2H4} on the gold cluster cations \ce{Au_{n}^{+}}
is very different from that of \ce{Au_{n}^{-}}. 
Thus, \ce{C2H4} readily adsorbs on the cationic 
monomer \ce{Au^+} and dimer \ce{Au_{2}^{+}} with an adsorption energy of
2.95 eV and 3.58 eV, respectively. Further increase in cluster size results in 
the sharp decrease in  $E_{\rm ads}$ up to 2.06 eV for $n=$3 followed by 
its slow decrease for 3$\le n \le$6.  At $n=$7 the adsorption energy $E_{\rm ads}$
exhibits a maximum followed by the second sharp decrease at $n=$ 8 and 9. 

For any $n$ in the range 1$\le n \le$10, the adsorption energy of 
the $\pi$-bonded \ce{C2H4}  on 
cationic $E_{\rm ads}^{\pi}(Au_n^+)$, 
neutral, $E_{\rm ads}^{\pi}(Au_n)$, 
and 
anionic, $E_{\rm ads}^{\pi}(Au_n^-)$,
gold clusters satisfies the relation:
$E_{\rm ads}^{\pi}(Au_n^+) > E_{\rm ads}^{\pi}(Au_n) >  E_{\rm ads}^{\pi}(Au_n^-)$.

Indeed, the excess of the positive charge 
on the cluster favours the process of electron transfer from the \ce{C2H4} 
to the gold cluster, which stabilizes the $\pi$ mode of adsorption, 
and strengthens the Au--C bond. 
A similar effect has been found for adsorption of \ce{CO} on small 
anionic and cationic gold clusters\cite{Wu02,Lee94}.
It was found that the adsorption energy of \ce{CO} 
is larger in the case of a cluster of cations and smaller in the case of anions. 
With the increasing cluster size, the differences in adsorption energies for 
different charge states become smaller\cite{Wu02}.

It is important to mention that 
strengthening the Au--C bond for \ce{C2H4} adsorbed on the gold clusters 
with an excess of the positive charge
is responsible for a cooperative effect in simultaneous adsorption
of \ce{C2H4} and \ce{O2} on neutral gold clusters.\cite{Lyalin09} 
Indeed, the adsorption of \ce{O2} on \ce{Au_n} 
results in an
electron transfer from the gold cluster to the 
anti-bonding $2\pi^*$ orbital of \ce{O2}. In this case, \ce{C2H4} effectively 
adsorbs on the positively charged gold cluster. 
A similar effect has also been predicted for adsorption of propene on \ce{O2-Au_n} 
clusters.\cite{Metiu04}
It was supposed that the binding of propene to \ce{O2-Au_n} should be stronger 
than the binding to the \ce{Au_n} cluster due to the effective charge transfer from the 
gold cluster to the oxygen molecule.\cite{Metiu04}

The excess of the positive or the negative charge on gold clusters 
results in a considerable change in the adsorption 
energy of the di-$\sigma$-bonded \ce{C2H4}.
As has been discussed above, the adsorption energy of the 
di-$\sigma$-bonded \ce{C2H4} on small neutral gold clusters 
exhibit an odd-even oscillations as a function of the cluster size.
These oscillations appear as a result of the electronic shell 
effects in gold clusters: \ce{Au_n} with odd $n$ have one uncoupled electron
that can easily transfer to the empty $\pi^*$ anti-bonding orbital of \ce{C2H4}. 
Such an electron transfer favors the di-$\sigma$ configuration of the adsorbed \ce{C2H4}.

\ref{fig:ads_minus_plus}a demonstrates  that 
there is no odd-even oscillations in the adsorption energy  
of the di-$\sigma$-bonded \ce{C2H4} on the anionic \ce{Au_{n}^{-}} clusters.
In that case, the cluster always possesses a weakly bounded electron that it can transfer to
the adsorbate.

In the case of the cationic \ce{Au_{n}^{+}} clusters, 
the electron back-donation process to \ce{C2H4} is suppressed for any $n$.
\ref{fig:ads_minus_plus}b shows that  $E_{\rm ads}$ calculated for
the  di-$\sigma$-bonded \ce{C2H4} on \ce{Au_{n}^{+}} exhibit two profound 
minima at $n=$3 and 9. 
In accordance with the spherical jellium model, 
\ce{Au_{3}^{+}}  and \ce{Au_{9}^{+}} clusters possess closed shell 
electronic structure with  2 and 8 valence electrons, respectively.
Hence, \ce{Au_{3}^{+}}  and \ce{Au_{9}^{+}} are chemically inactive, 
in particular in the processes where electron transfer from the cluster to the 
adsorbate is responsible for formation of the chemical bond.

\ref{fig:ads_minus_plus}c and \ref{fig:ads_minus_plus}d  demonstrate
the $n$-evolution of the C--C bond distance, $r_{C-C}$, in
\ce{C2H4} adsorbed on \ce{Au_{n}^{-}} and \ce{Au_{n}^{+}} clusters, respectively.
The C--C bond in the $\pi$-bonded \ce{C2H4}  on the anionic and the cationic clusters
is enlarged up to 1.36--1.40 \AA\, irrespective to cluster charge. 
This is 0.04 - 0.08 \AA\ larger than that for the free \ce{C2H4} (1.32 \AA). 
The di-$\sigma$-bonded ethylene is activated strongly in the case of adsorption on 
the anionic \ce{Au_{n}^{-}} clusters.
Thus, in the di-$\sigma$-bonded ethylene the C--C bond 
is enlarged up to 1.53--1.42 \AA\ for the anionic and only up to 1.48--1.37 \AA\ for the 
cationic gold clusters. Such an effect has a clear explanation. 
The excess of electrons on a cluster promotes back-donation of an electron 
from the cluster to the $\pi^*$ anti-bonding orbital of \ce{C2H4}, hence promoting its 
activation. The excess of the positive charge on a cluster suppresses the back-donation process, thereby suppressing the \ce{C2H4} activation.
Thus, by changing the cluster charge one can manipulate the cluster reactivity. 

\subsection{Vibrational spectra of ethylene adsorbed on small gold clusters}

Analysis of the vibrational spectrum of adsorbed \ce{C2H4} directly reveals 
the mode of \ce{C2H4} adsorption and the state of hybridization. This is due to the
sensitivity of the C--C stretching frequency to the hybridized state of carbon 
atoms.\cite{Stuve85}
In 1985 Stuve and Madix introduced the so-called $\pi\sigma$ parameter
for characterizing rehybridization in \ce{C2H4} upon adsorption 
on metal surfaces.\cite{Stuve85}  
The $\pi\sigma$ parameter depends on the 
shift of the C--C stretching frequencies in \ce{C2H4} and ranges from 
0 for the free \ce{C2H4}, to 0.38 for \ce{K[(C2H4)PtCl3]} (Zeise's salt), 
and finally to 1 for \ce{C2H4Br2}. 
Zeise's salt and \ce{C2H4Br2} have been chosen by Stuve and Madix as models describing 
pure $\pi$- and di-$\sigma$-bonding of \ce{C2H4}, respectively.
Thus the larger the shift in vibrational frequencies 
(compared to the free \ce{C2H4}), the greater the degree of rehybridization.

A similar approach can be applied for \ce{C2H4} adsorbed on metal clusters. 
\ref{tbl:freq} presents the fundamental frequencies of free \ce{C2H4} calculated in 
the harmonic approximation and obtained from experiment\cite{Shimanouchi72}.  

\begin{table}
  \caption{Fundamental frequencies of \ce{C2H4} }
  \label{tbl:freq}
  \begin{tabular}{lcc}
    \hline
    Mode & Harmonic, cm$^{-1}$ & Experiment,\cite{Shimanouchi72} cm$^{-1}$\\
    \hline
    $\nu_{1}$  $(b_{2u})$      & 826  & 826 \\
    $\nu_{2}$  $(b_{3u})$      & 975  & 943 \\
    $\nu_{3}$  $(b_{2g})$      & 981  & 949 \\
    $\nu_{4}$  $(a_{u})$       & 1057 & 1023 \\
    $\nu_{5}$  $(b_{3g})$      & 1237 & 1236 \\
    $\nu_{6}$  $(a_{g})$       & 1377 & 1342 \\
    $\nu_{7}$  $(b_{1u})$      & 1467 & 1444 \\
    $\nu_{8}$  $(a_{g})$       & 1693 & 1623 \\
    $\nu_{9}$  $(b_{1u})$      & 3131 & 2989 \\
    $\nu_{10}$ $(a_{g}) $      & 3146 & 3026 \\
    $\nu_{11}$ $(b_{3g})$      & 3205 & 3103 \\
    $\nu_{12}$ $(b_{2u})$      & 3233 & 3106 \\
    \hline
  \end{tabular}
\end{table}

The obtained theoretical results are in a good agreement with experimental data.
Some discrepancy, especially noticeable 
in the high frequency range, might result from anharmonic effects. 
The anharmonic corrections can be taken into account, for example by the method
of n-mode coupling representations of the quadratic force field.\cite{Yagi04}
However such a consideration goes far beyond the scope 
of the present paper. In order to distinguish between $\pi$ and di-$\sigma$ modes of 
\ce{C2H4} adsorption one should to calculate the shift in the C--C stretching 
frequencies of the adsorbed \ce{C2H4} with respect to the free molecule. In that case the 
anharmonic corrections to the vibrational energy levels can be partly canceled. Therefore in this
paper, we perform analysis of the vibrational spectra of the adsorbed \ce{C2H4} 
in the harmonic approximation.

\begin{figure}[htbp]
\includegraphics[scale=1.25,clip]{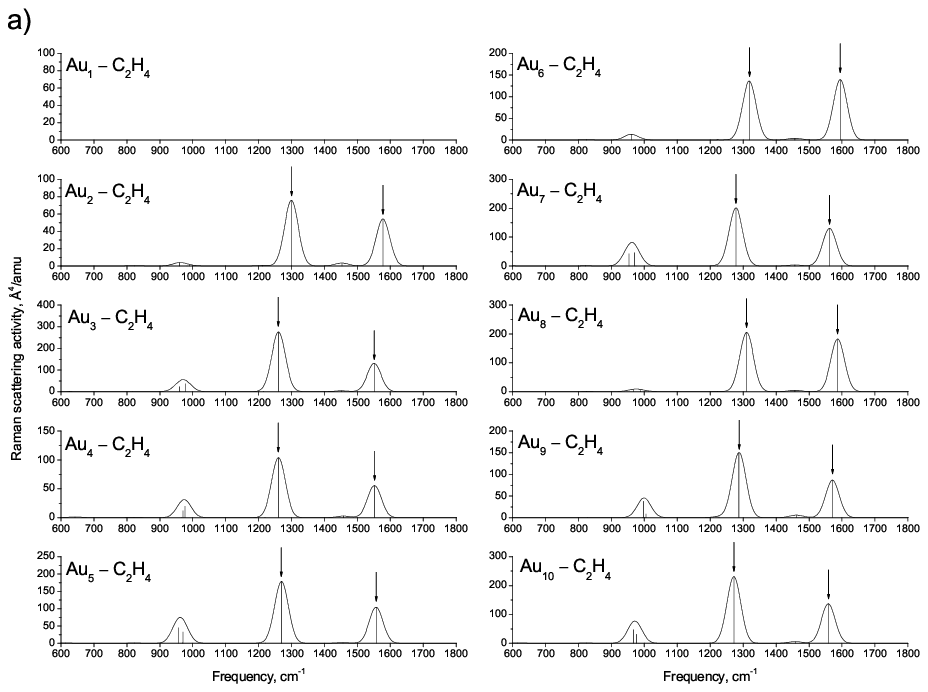}
\includegraphics[scale=1.25,clip]{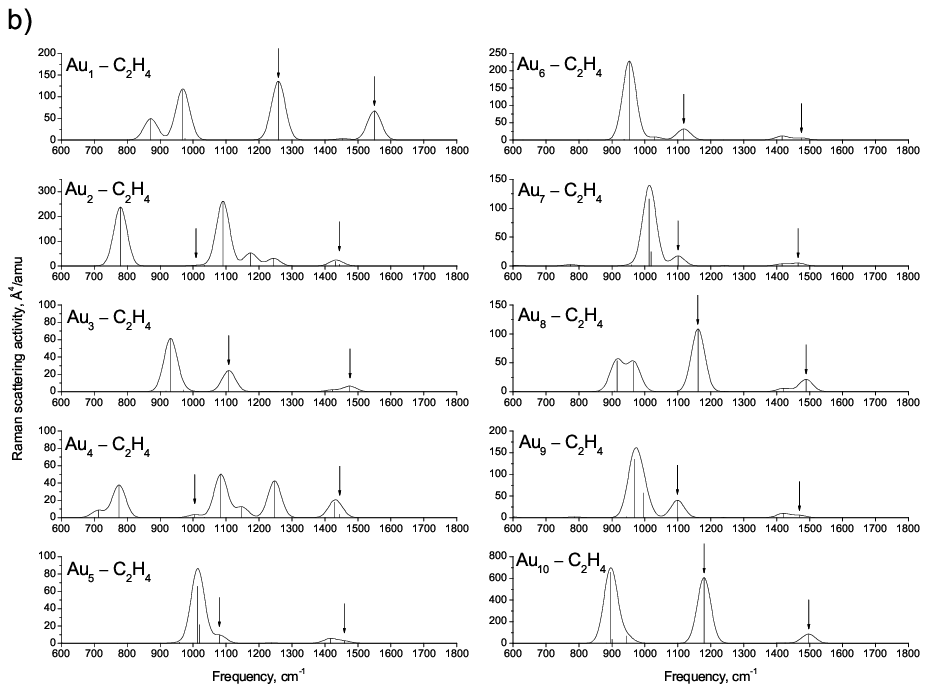}
\caption{Raman scattering activity of the neutral \ce{C2H4-Au_n} clusters:
(a) $\pi$-bonded \ce{C2H4};
(b) di-$\sigma$-bonded \ce{C2H4}. 
Gaussian broadening of half-width 45 cm$^{-1}$ has been used.
The stretching modes corresponding to the $\nu_6$ and $\nu_8$ vibrations in \ce{C2H4} 
are marked by the vertical arrows. Note, that in the case of a 
single Au atom, only one $\sigma$ bond is formed.}
\label{fig:freq_neutral}
\end{figure} 

Following Ref.\cite{Stuve85}, we have selected two vibrational modes: 
$\nu_6$ and $\nu_8$, corresponding to the direct C--C and the in-plane \ce{CH2} 
scissoring modes of vibrations, respectively. 
Information on the shift of the vibrational frequencies of \ce{C2H4} upon its adsorption 
can be obtained from infrared or Raman spectroscopy.
\ref{fig:freq_neutral} presents the frequency dependence 
of the Raman scattering activity calculated for a 
\ce{C2H4} molecule adsorbed in the $\pi$ (\ref{fig:freq_neutral}a) and di-$\sigma$ (\ref{fig:freq_neutral}b) configurations 
on the neutral \ce{Au_n} clusters with $n$ = 1, ..., 10.   
The stretching modes corresponding to the $\nu_6$ and $\nu_8$ vibrations in \ce{C2H4} 
are marked by the vertical arrows. 
It is seen from \ref{fig:freq_neutral}a that the spectral dependences of the  
Raman scattering activity calculated for the $\pi$-bonded ethylene on \ce{Au_n} clusters 
are  similar for all the cluster sizes $n$ considered. 
The first maximum at ~ 960 cm$^{-1}$ corresponds to the symmetric and asymmetric 
swinging of H atoms in the direction perpendicular to the plane of \ce{C2H4}--- 
similar to the $\nu_2$ and $\nu_3$ vibrational modes in free \ce{C2H4}.
These vibrations remain unchanged in the $\pi$-bonded \ce{C2H4}. 
Two intensive maxima at ~ 1280 cm$^{-1}$ and ~ 1560 cm$^{-1}$
correspond to the $\nu_6$ and $\nu_8$ vibrational modes in \ce{C2H4}. These lines 
are red-shifted in comparison with the free \ce{C2H4}.

\begin{figure}[htb]
\includegraphics[scale=1.2,clip]{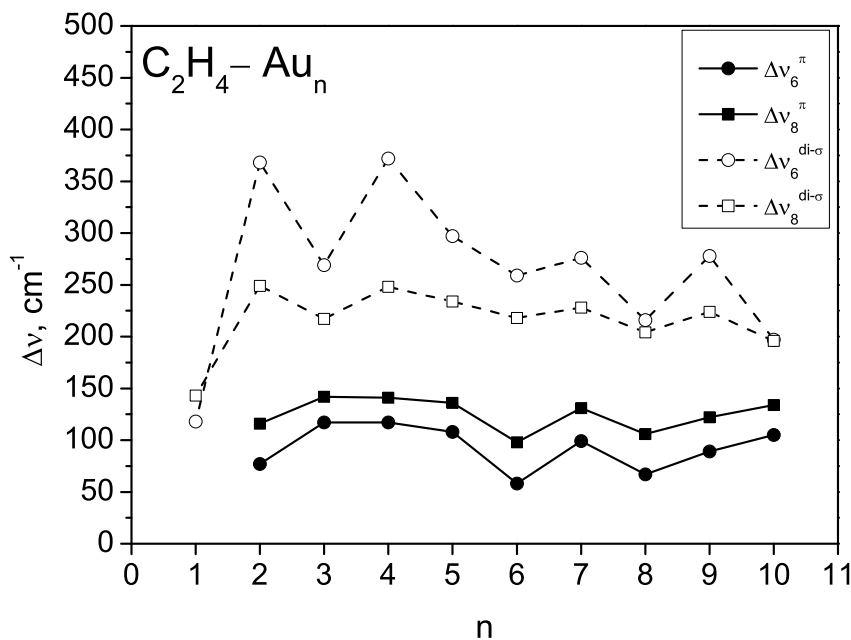}
\caption{Shift in the harmonic frequencies, $\nu_6$ and $\nu_8$,  
calculated for the $\pi$ (solid lines) and di-$\sigma$ (dashed lines) configurations 
of adsorbed \ce{C2H4} on the neutral \ce{Au_n} clusters with $1 \le n \le 10$.
Note, that in the case of a single Au atom, only one $\sigma$ bond is formed.}
\label{fig:shift}
\end{figure} 

Solid lines in \ref{fig:shift} demonstrate shifts 
$\Delta \nu_6 ^{\pi}$ and $\Delta \nu_8 ^{\pi}$ in harmonic frequencies $\nu_6$ and $\nu_8$
due to the $\pi$ mode of \ce{C2H4} adsorption on small neutral gold clusters. 
For the $\pi$-bonded \ce{C2H4}, the scissoring mode $\nu_8$ is shifted by 30-40 cm$^{-1}$ 
towards low frequencies if compared with the $\nu_6$ mode. 
Both the $\Delta \nu_6 ^{\pi}$ and $\Delta \nu_8 ^{\pi}$ shifts
vary in the range of 58 - 142 cm$^{-1}$ and exhibit similar dependence 
on the cluster size with local minima at $n$=6 and 8.  The relatively small frequency shifts 
$\Delta \nu_6 ^{\pi}$ and $\Delta \nu_8 ^{\pi}$ correspond to the values of the $\pi\sigma$ 
parameter typical for the $\pi$-bonded \ce{C2H4}.

\ref{fig:freq_neutral} demonstrates that the spectral behavior 
of the Raman scattering activity calculated for 
the di-$\sigma$-bonded \ce{C2H4} is different from that obtained for 
the $\pi$-bonded \ce{C2H4}.  
Thus, for cluster sizes $n$= 2, 4, 5, 7 and 9,
a broad band appears in the frequency region of 900 - 1300 cm$^{-1}$.
This band results from the strong mixing of the C--C stretching mode $\nu_6$
with the hydrogen swinging ($\nu_2$ and $\nu_3$) and twisting ($\nu_4$) modes.  
The coupling of the  $\nu_2$, $\nu_3$, $\nu_4$, and $\nu_6$ vibrational modes occurs
as a result of considerable deformations in \ce{C2H4}. These deformations include
the bending of H atoms in respect to the plane of a free \ce{C2H4}
due to the $sp^3$ hybridization in di-$\sigma$ bonded 
\ce{C2H4}, as well as the twisting of \ce{CH2} groups in respect to each other along the C--C bond.
Such twisting is 
especially noticeable for \ce{C2H4-Au2} and \ce{C2H4-Au4} clusters. 
The strong coupling between different vibrational modes makes identification 
of $\nu_6$ and $\nu_8$ vibrations to be rather difficult: the mixed modes 
form new levels, neither of which represent the true  $\nu_6$ and $\nu_8$ states.

Appearance of the $sp^3$ hybridization character in di-$\sigma$ bonded 
\ce{C2H4} results in a considerable red-shift in the stretching frequencies. 
Such shifts $\Delta \nu_6 ^{{\rm di-}\sigma}$ and $\Delta \nu_8 ^{{\rm di-}\sigma}$ 
in harmonic frequencies $\nu_6$ and $\nu_8$ are shown by the dashed lines in \ref{fig:shift}. 

It is seen that the $\Delta \nu_6 ^{{\rm di-}\sigma}$ shift is considerably larger than the 
$\Delta \nu_8 ^{{\rm di-}\sigma}$ which can be explained by the strong coupling between the 
$\nu_2$, $\nu_3$, $\nu_4$, and $\nu_6$ vibrational levels. 
The $\Delta \nu_6 ^{{\rm di-}\sigma}$ and $\Delta \nu_8 ^{{\rm di-}\sigma}$ frequency 
shifts varies in the range of 200 - 370 cm$^{-1}$ for $n$=2, ... ,10. 
These values correspond to the $\pi\sigma$ 
parameter typical for the di-$\sigma$-bonded \ce{C2H4}. Thus, 
the analysis of the vibrational modes of \ce{C2H4} adsorbed on gold clusters 
allows one to identify the specific modes of \ce{C2H4} adsorption and the 
rate of rehybridization upon adsorption.

\section{Conclusions}

In summary, we have demonstrated that \ce{C2H4} can be adsorbed
on small gold clusters in two different configurations,
corresponding to the $\pi$- and di-$\sigma$-bonded species.
Adsorption in the $\pi$-bonded mode dominates over the di-$\sigma$ 
mode over all considered cluster sizes $n$, with the exception of the 
neutral \ce{C2H4-Au5} system. We found a striking difference in 
the size dependence of the 
adsorption energy of \ce{C2H4} bonded to neutral gold clusters in the 
$\pi$ and di-$\sigma$ configurations.  The adsorption energy, calculated 
for the di-$\sigma$-bonded \ce{C2H4} exhibits 
pronounced odd-even oscillations, showing the importance of the electronic 
shell effects in the di-$\sigma$ mode of ethylene adsorption on gold clusters. 
The different rate of hybridization in the $\pi$- and di-$\sigma$-bonded 
\ce{C2H4} can be responsible for the different rate of catalytic activation 
and hence, different reactivity of adsorbed \ce{C2H4}.

We have also demonstrated that the interaction of \ce{C2H4} with small gold clusters 
strongly depends on the cluster charge. Hence ethylene adsorption and reactivity  
can be manipulated by the cluster charge.
The strengthening of the ethylene -- cluster interaction for gold clusters 
with an excess of the positive charge is responsible for the cooperative effect 
in simultaneous adsorption of \ce{C2H4} and \ce{O2} on neutral gold clusters. 
This effect can play an important role in the mechanism of
catalytic oxidation of alkenes by dioxygen on the surface of gold
clusters.

Finally we have shown that the analysis of the vibrational modes of \ce{C2H4} 
adsorbed on gold clusters allows one to identify the specific modes 
of \ce{C2H4} adsorption, the rate of rehybridization upon adsorption, and thus the
\ce{C2H4} reactivity.

In the present work, we have considered adsorption of ethylene molecule
onto small gold clusters consisting of up to 10 atoms. 
Many interesting problems beyond the scope
of the present work arise when considering multiple molecular 
adsorption on gold nanoparticles of larger
sizes up to 1-5 nm, where the strong dependence of the catalytic
activity of gold nanoparticles has been observed experimentally.
In particular, multiple molecular adsorption on the surface of
gold nanoparticles can result in a considerable change in 
cluster morphology, even for relatively large cluster sizes. 

Another important direction for future development is to
investigate the effect of alloying on the adsorption and catalytic activation 
of alkenes on gold clusters. The doping of a gold cluster with atomic impurity 
can result in considerable changes in the geometrical and electronic structure of the cluster, 
thereby modifying and controlling the catalytic activity. 
Understanding how to enhance and control chemical reactions on a cluster surface 
is therefore a vital task for nanocatalysis.
 
\begin{acknowledgement}
This work was supported by the Global COE Program
(Project No. B01:  Catalysis as the Basis for Innovation 
in Materials Science) from the Ministry of Education, 
Culture, Sports, Science and Technology, Japan.
Calculations were performed in the Research Center for Computational Sciences,
Okazaki Research Facilities.
\end{acknowledgement}

\end{document}